\documentclass[fleqn,12pt,twoside]{article}
\usepackage{espcrc1}
\usepackage{amsmath}
\usepackage{amssymb}
\usepackage{graphicx}
\title{Abelian and Center Vortex Condensation in
SU(3) Lattice Gauge Theory}

\author{Paolo Cea\address[DF,INFN]{Dipartimento di Fisica, Univ. of Bari and INFN - Sezione di Bari,
        I-70126 Bari, Italy}  and
        Leonardo Cosmai\address[INFN]{INFN - Sezione di Bari, I-70126 Bari, Italy}}

\begin{document}

\maketitle

\begin{abstract}
 We study the condensation of Abelian and Center vortices in
 SU(3) lattice gauge theory at finite
 temperature.
 We find that  both vortices condense in the confined phase
 of the SU(3) vacuum.
\end{abstract}

\section*{ }%

To study the vacuum structure of the lattice gauge theories we
introduced~\cite{Cea:1997ff,Cea:1999gn} a gauge invariant
effective action for the external static
background field
$\vec{A}^{\mathrm{ext}}(\vec{x})$
\begin{equation}
\label{Gamma}
 \Gamma[\vec{A}^{\mathrm{ext}}] =
-\frac{1}{L_4} \ln \left\{
\frac{{\mathcal{Z}}[\vec{A}^{\mathrm{ext}}]}{{\mathcal{Z}}[0]}
\right\} \,\,,\,\,\,\, \mathcal{Z}[U^{\mathrm{ext}}_\mu] =
\int_{U_k(x)|_{x_4=0} = U^{\mathrm{ext}}_k(x)} {\mathcal{D}}U \;
e^{-S_W} \,.
\end{equation}
$\mathcal{Z}[U^{\mathrm{ext}}_\mu]$ is the lattice Schr\"odinger
functional (invariant, by definition, for lattice gauge
transformations of the external links), $U^{\mathrm{ext}}_k(x)$ is
the lattice version of the external continuum gauge field
$\vec{A}^{\mathrm{ext}}(x)= \vec{A}^{\mathrm{ext}}_a(x)
\lambda_a/2$, and $S_W$ is the standard Wilson action.
$\mathcal{Z}[0]$ is the lattice Schr\"odinger functional with
$\vec{A}^{\mathrm{ext}}=0$ ($U^{\mathrm{ext}}_\mu={\mathbf{1}}$).

At finite temperature  we introduced the thermal partition
function in presence of a given static background field:
\begin{equation}
\label{ZetaTnew} \mathcal{Z}_T \left[ \vec{A}^{\text{ext}} \right]
=
\int_{U_k(\beta_T,\vec{x})=U_k(0,\vec{x})=U^{\text{ext}}_k(\vec{x})}
\mathcal{D}U \, e^{-S_W}   \,, \,\,  \beta_t=L_4=\frac{1}{a T} \,.
\end{equation}
If we send the physical temperature to zero the thermal functional
Eq.~(\ref{ZetaTnew}) reduces to the zero-temperature Schr\"odinger
functional given in Eq.~(\ref{Gamma}).

We would like to detect vortex condensation. To this purpose we
use our lattice effective action to define a disorder
parameter~\cite{DiGiacomo:1999fa,DiGiacomo:1999fb} with non
vanishing vacuum expectation value in the confined phase:
\begin{equation}
\label{disorderT}
\mu = e^{-F_{\text{vort}}/T_{\text{phys}}} =
\frac{\mathcal{Z}_T[\text{vort}]}{\mathcal{Z}_T[0]} \,,
\end{equation}
where $\mathcal{Z}_T[{\text{vort}}]$ and $\mathcal{Z}_T[0]$ are,
respectively,  the thermal partition function with a vortex
background field and without the background field,
$F_{\text{vort}}$ is the free energy to create a vortex.

In SU(3) gauge theory we may consider two independent kinds of
Abelian vortices $T_3$ and $T_8$ (and their linear combinations)
associated respectively to the generators $\lambda_3$ and
$\lambda_8$ (for details see~\cite{Cea:2000zr}).

In case of center vortices the thermal partition function
$\mathcal{Z}_T[{\mathcal{P}_{\mu \nu}}]$ is defined by multiplying
by the center element $\exp(i 2 \pi/3)$ the set $\mathcal{P}_{\mu
\nu}$ of plaquettes~\cite{Kovacs:2000sy,DelDebbio:2000cx}
$\mathcal{P}_{\mu \nu}(x_1,x_2,x_3,x_4)$ with $(\mu,\nu)=(4,2)$,
$x_4=x_4^\star$, $x_2=\frac{L_s}{2}$ and $L_s^{\text{min}} \le
x_{1,3} \le L_s^{\text{max}}$ , with $L_s$ the lattice spatial
linear size.

\begin{figure}[htb]
\begin{minipage}[t]{80mm}
\includegraphics[width=1.0\textwidth,clip]{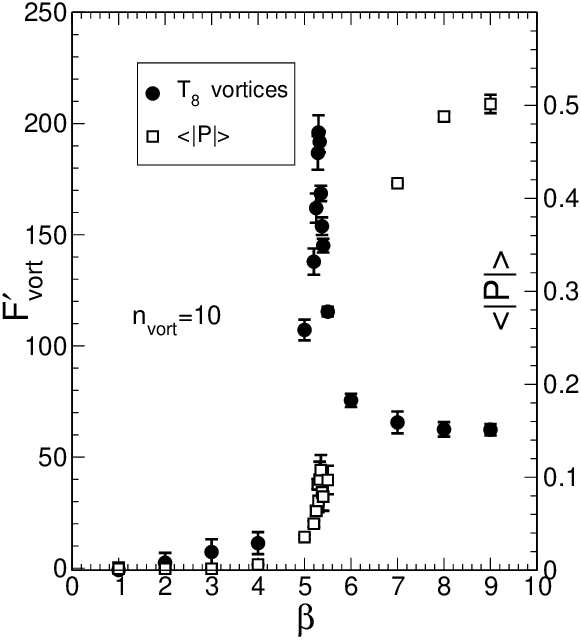}
\vspace{-1.5cm} \caption{$F^{\prime}_{\text{vort}}$, the
$\beta$-derivative of the free energy $F_{\text{vort}}$ for $T_8$
Abelian vortices display a peak in correspondence of the rise of
the absolute value of the Polyakov loop.}
\label{fig:1}
\end{minipage}
\hspace{\fill}
\begin{minipage}[t]{76mm}
\includegraphics[width=1.0\textwidth,clip]{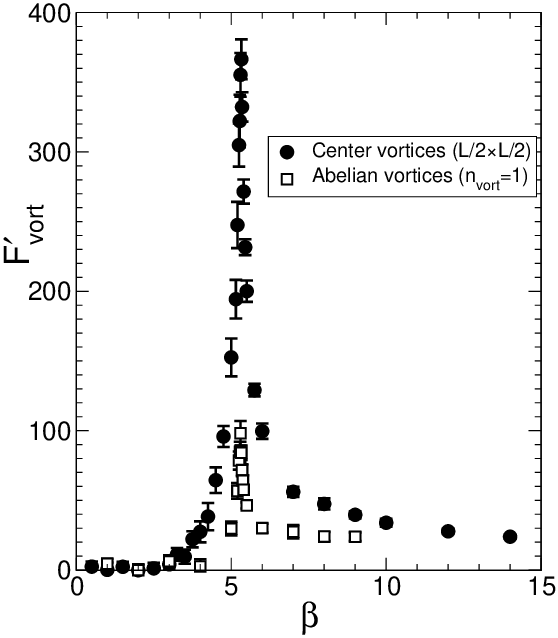}
\vspace{-1.5cm} \caption{$F^{\prime}_{\text{vort}}$ for center
vortices ($L_s^{\text{min}}=L/2$) and $T_8$ Abelian vortices
(vortex charge $n_{\text{vort}}=1$).} \label{fig:2}
\end{minipage}
\end{figure}
\vspace{-0.3cm} By numerical integration of
$F^{\prime}_{\text{vort}}$ we can compute $F_{\text{vort}}$ and
the disorder parameter $\mu$ (see Eq.~(\ref{disorderT})). Our
numerical results suggest that in the confined phase of SU(3)
lattice gauge theory $F_{\text{vort}}=0$ (in the thermodynamic
limit).


\end{document}